\documentclass[final]{sig-alternate-05-2015}

\usepackage{url}           
\usepackage{listings}      
\usepackage{enumitem}      
\usepackage[colorlinks=true,allcolors=blue,breaklinks,draft=false]{hyperref}   
\usepackage{graphicx}
\usepackage{float}
\usepackage{subcaption}
\usepackage{xspace,framed}
\usepackage{colortbl}
\usepackage{calc}
\usepackage[dvipsnames]{xcolor}
\usepackage{amssymb,amsmath,amsfonts} 
\usepackage{mathtools}
\usepackage{microtype}

\usepackage{algorithm}
\usepackage{amsfonts}
\usepackage{multicol}
\usepackage{multirow}
\usepackage{tikz}
\usetikzlibrary{positioning, automata, shapes.arrows, calc, shapes, arrows}
\usetikzlibrary{patterns}
\usepackage[justification=centering]{caption}
\usepackage{stmaryrd}
\usepackage{hhline}
\usepackage{pifont}
\usepackage{longtable}
\usepackage{afterpage}
\usepackage{wasysym}
\usepackage[scaled]{helvet}
\usepackage{mdwlist}

\newcommand{\aabatecmt}[1]{}
\newcommand{\dariocmt}[1]{}

\newtheorem{myassumption}{Assumption}
\newtheorem{mylemma}{Lemma}
\newtheorem{myprop}{Proposition}

\allowdisplaybreaks

\newcommand{\param}[2]{\ensuremath{\langle{#1},{#2}\rangle}\xspace}
\newcommand{\xmark}{\ding{55}}

\begin{document}

\setcopyright{acmcopyright}
\isbn{}
\doi{}

\title{
Sound and Automated Synthesis of Digital Stabilizing Controllers for Continuous
Plants%
\thanks{Supported by EPSRC grant EP/J012564/1,
ERC project 280053 (CPROVER) and the H2020 FET OPEN SC$^2$.}}

\author{Alessandro Abate$^{1}$, Iury Bessa$^{2}$, Dario Cattaruzza$^{1}$, Lucas Cordeiro$^{1,2}$, \\ 
Cristina David$^{1}$, Pascal Kesseli$^{1}$ and Daniel Kroening$^{1}$
\and
$^{1}$\affaddr{University of Oxford, Oxford, United Kingdom} \quad
$^{2}$\affaddr{Federal University of Amazonas, Manaus, Brazil}
}

\newcommand\tool{{\sf DSSynth}\xspace}

\maketitle

\begin{abstract}
Modern control is implemented with digital microcontrollers, embedded within
a dynamical plant that represents physical components.
We present a new algorithm based on counter\-example guided inductive
synthesis that automates the design of digital controllers that are
correct by construction.  The synthesis result is sound with respect to the
complete range of approximations, including time discretization,
quantization effects, and finite-precision arithmetic and its
rounding errors.
We have implemented our new algorithm in a tool called \tool, and are able
to automatically generate stable controllers for a set of intricate plant
models taken from the literature within minutes.
\end{abstract}

%
%
\printccsdesc

\keywords{Digital control synthesis, CEGIS, finite-word-length representation, time sampling, quantization}

\section{Introduction}

Modern implementations of embedded control systems have proliferated with
the availability of low-cost devices that can perform highly non-trivial
control tasks, with significant impact in numerous application areas such as
environmental control and robotics~\cite{astrom1997computer, Franklin15}. 
Correct control is non-trivial, however.  The problem is exacerbated by
artifacts specific to digital control, such as the effects of
finite-precision arithmetic, time discretization, and quantization noise
introduced by A/D and D/A conversion.  Thus, programming expertise is a key
barrier to broad adoption of correct digital controllers, and requires
considerable knowledge outside of the expertise of many control engineers.

Beyond classical a-posteriori validation in digital control, there has been
plenty of previous work aiming at \emph{verifying} a given designed
controller, which however broadly lack automation.  Recent work has studied
the stability of digital controllers considering implementation aspects,
i.e., fixed-point arithmetic and the word length~\cite{Bessa16}.  They
exploit advances in bit-accurate verification of C programs to obtain a
verifier for software-implemented digital control.

By contrast, we leverage a very recent step-change in the automation and
scalability of \emph{program synthesis}.  Program synthesis engines use a
specification as the starting point, and subsequently generate a sequence of
candidate programs from a given template.  The candidate programs are
iteratively refined to eventually satisfy the specification.  Program
synthesizers implementing Counter-Example Guided Inductive Synthesis
(CEGIS)~\cite{sketch} are now able to generate programs for highly
non-trivial specifications with a very high degree of automation.  Modern
synthesis engines combine automated testing, genetic algorithms, and
SMT-based automated reasoning~\cite{DBLP:journals/corr/AlurFSS16a,
DBLP:conf/lpar/DavidKL15}.

By combining and applying state-of-the-art synthesis engines we present a
tool that automatically generates digital controllers for a given continuous
plant model that are correct by construction.  This approach delivers a high
degree of automation, promises to reduce the cost and time of development of
digital control dramatically, and requires considerably less expertise than
a-posteriori verification.  Specifically, we synthesize stable,
software-implemented embedded controllers along with a model of a physical
plant.  Due to the complexity of such closed-loop systems, in this work we
focus on linear models with known configurations, and perform parametric
synthesis of stabilizing digital controllers (further closed-loop performance
requirements are left to future work).

Our work addresses challenging aspects of the control synthesis problem.  We
perform digital control synthesis over a hybrid model, where the plant
exhibits continuous behavior whereas the controller operates in discrete
time and over a quantized domain.  Inspired by a classical
approach~\cite{astrom1997computer}, we translate the problem into a single
digital domain, i.e., we model a digital equivalent of the continuous plant
by evaluating the effects of the quantizers (A/D and D/A converters) and of
time discretization.  We further account for uncertainties in the plant
model.  The resulting closed-loop system is a program with a loop that
operates on bit-vectors encoded using fixed-point arithmetic with finite
word length (FWL).  The three effects of 1.~uncertainties, 2.~FWL
representation and 3.~quantization errors are incorporated into the model,
and are taken into account during the CEGIS-based synthesis of the control
software for the plant.

In summary, this paper makes the following original contributions.
\begin{itemize*}

\item We automatically generate {\em correct-by-construction} digital
  controllers using an inductive synthesis approach.  Our application of program
  synthesis is non-trivial and addresses challenges specific to control
  systems, such as the effects of quantizers and FWL.  In particular, we
  have found that a two-stage verification engine that continuously refines
  the precision of the fixed-point representation of the plant yields a
  speed-up of two orders of magnitude over a conventional one-stage
  verification engine.

\item Experimental results show that \tool is able to efficiently synthesize
  stable controllers for a set of intricate benchmarks taken from the
  literature: the median runtime for our benchmark set considering the
  faster engine is $48$\,s, i.e., half of the controllers can be synthesized
  in less than one minute.



\end{itemize*}

\section{Preliminaries}\label{sec:preliminaries}

\subsection{Discretization of the Plant}
\label{ssec:SandH}

The digital controllers synthesized using the algorithm we present in this
paper are typically used in closed loops with continuous (physical) plants. 
Thus, we consider continuous dynamics (the plant) and discrete parts (the
digital controller). In order to obtain an overall model for the synthesis, 
we discretize the continuous plant and particularly look at the plant dynamics 
from the perspective of the digital controller. 

As we only consider transfer function models, and require a $z$-domain transfer
function $G(z)$ that captures all aspects of the continuous plant, 
which is naturally described via a Laplace-domain transfer function $G(s)$. 
The continuous model of the plant must be discretized to obtain the
corresponding coefficients of $G(z)$.

Among the discretization methods in the literature~\cite{Franklin15}, we
consider the sample-and-hold processes in complex
systems~\cite{istepanian2012digital}.  On the other hand, the ZOH
discretization models the exact effect of sampling and DAC interpolation
over the plant.

\begin{myassumption}
The sample-and-hold effects of the ADC and the presence of the ZOH of the DAC are synchronized,
namely there is no delay between sampling the plant output at the ADC and
updating the DAC accordingly.  The DAC interpolator is an ideal ZOH. 
\end{myassumption}

\begin{mylemma}\cite{astrom1997computer}
Given a synchronized ZOH input and sample-and-hold output on the plant, 
with a sample time $T$ satisfying the Nyquist criterion, the discrete pulse
transfer function $G(z,T)$ is an exact z-domain representation of $G(s)$, 
and can be computed using the following formula:
\begin{equation}
\label{eq:pulsetf}
G(z,T) = 
(1-z^{-1})\mathcal{Z}\left\lbrace{\mathcal{L}^{-1}\left\lbrace{\frac{G(s)}{s}}\right\rbrace_{t=kT}}\right\rbrace.
\end{equation}
\end{mylemma}
In this study, for the sake of brevity, we will use the notation $G(z)$ to represent the
pulse transfer function $G(z,T)$.
Lemma~\ref{eq:pulsetf} ensures that the poles and zeros match under the
$\mathcal{Z}\left\lbrace{\mathcal{L}^{-1}\left\lbrace{\cdot}\right\rbrace_{t=kT}}\right\rbrace$
operations, and it includes the ZOH dynamics in the $(1-z^{-1})$ term.  This
is sufficient for stability studies over $G(s)$~\cite{fadali}, i.e., if
there is any unstable pole (in the complex domain $\Re\{s\} > 0$), the pulse
transfer function in~\eqref{eq:pulsetf} will also present the same number of
unstable poles ($|z| > 1$)~\cite{Franklin15}.

\subsection{Model Imprecision, Finite Word Length Representation and Quantization Effects}
\label{verifying-closed-loop-control-systems}

Let $C(z)$ be a digital controller and $G(z)$ be a discrete-time representation of the plant, 
given as
\begin{align}
\small
\label{controller_plant_tf}
C(z)&=\frac{C_n(z)}{C_d(z)}=\frac{\beta_{0}+\beta_{1}z^{-1}+...+\beta_{M_C}z^{-M_C}}{\alpha_{0}+\alpha_{1}z^{-1}+...+\alpha_{N_C}z^{-N_C}}, \\
G(z)&=\frac{G_n(z)}{G_d(z)}=\frac{b_{0}+b_{1}z^{-1}+...+b_{M_G}z^{-M_G}}{a_{0}+a_{1}z^{-1}+...+a_{N_G}z^{-N_G}}.
\end{align}
where $\vec{\beta}$ and $\vec{\alpha}$ are vectors containing the
controller's coefficients; similarly, $\vec{b}$ and $\vec{a}$ denote the
plant's coefficients; 
and finally $N_{(\cdot)}$ and $M_{(\cdot)}$ indicate the order of the polynomials,  
and we require in particular that $N_G \geq M_G$. 

Uncertainties in $G(z)$ may appear owing to: 1.~uncertainties in $G(s)$ (we
denote the uncertain continuous plant by
$\hat{G}(s)=\frac{G_n(s)+\Delta_p{G}_n(s)}{G_d(s)+\Delta_p{G}_d(s)}$ to
explicitly encompass the effects of the uncertainty terms
$\Delta_p{G}_{(\cdot)}(s)$) arising from tolerances/imprecision in the
original model; 2.~errors in the numerical calculations due to FWL effects
(e.g., coefficient truncation and round-off, which will be denoted as
$\Delta_b{G}_n(s),\Delta_b{G}_d(s)$); and 3.~errors caused by quantization
(which we model later as as external disturbances $\nu_1$ and $\nu_2$). 
These uncertainties are parametrically expressed by additive terms,
eventually resulting in an uncertain model $\hat{G}(z)$, such that:
\begin{equation}
\hat{G}(z)=\frac{G_n(z)+\Delta G_n(z)}{G_d(z)+\Delta G_d(z)},
\end{equation}
%
which will be represented by the following transfer function:
\begin{equation}
\hat{G}(z)=\frac{\hat{b}_{0}+\hat{b}_{1}z^{-1}+...+\hat{b}_{M_G}z^{-M_G}}{\hat{a}_{0}+\hat{a}_{1}z^{-1}+...+\hat{a}_{N_G}z^{-N_G}}. 
\end{equation}
Notice that, due to the nature of the methods we use for the stability
check, we require that the parametric errors in the plant have the same
polynomial order as the plant itself (indeed, all other errors described in
this paper fulfill this property).  We~also remark that, due to its native
digital implementation, there are no parametric errors ($\Delta_pC_n(z),
\Delta_pC_d(z)$) in the controller.  Thus $\hat{C}(z) \equiv C(z)$.

We introduce next a notation based on the coefficients of the polynomial to
simplify the presentation.  Let $\mathcal{P}^{N}$ be the space of
polynomials of order $N$.  Let $P \in \mathcal{P}^{M,N}$ be a rational
polynomial $\frac{P_n}{P_d}$, where $P_n \in \mathcal{P}^{M}$ and $P_d \in
\mathcal{P}^{N}$.  For a vector of coefficients
\begin{equation}
\vec{P} \in \mathbb{R}^{N+M+2}=[n_{0}\ n_{1}\ \hdots \ n_{M}\ d_{0}\ d_{1}\ \hdots\ d_{N}\ ]^T
\label{eq:coefficients}
\end{equation}
and an uncertainty vector 
\begin{equation}
\Delta{\vec{P}}\in \mathbb{R}^{N+M+2}=[\Delta{n}_{0}\ \hdots \ \Delta{n}_{M}\ \Delta{d}_{0}\ \hdots\ \Delta{d}_{N}\ ]^T \; 
\label{eq:delta_coefficients}
\end{equation}
we write
\begin{align}
\label{eq:hatgvector}
\vec{\hat{G}}&=\vec{G}+\Delta{\vec{G}}, \text{ where} \\
\vec{G} \in \mathbb{R}^{N_G+M_G+2}&=[b_{0}\ \hdots \ b_{M_G}\ a_{0}\ \hdots\ a_{N_G}\ ]^T, \nonumber \\
\Delta{\vec{G}}\in \mathbb{R}^{N_G+M_G+2}&=[\Delta{b}_{0}\ \hdots \ \Delta{b}_{M_G}\ \Delta{a}_{0}\ \hdots\ \Delta{a}_{N_G}\ ]^T. \nonumber
\end{align}
In the following we will either manipulate the transfer functions $G(z)$, $C(z)$ directly, 
or work over their respective coefficients $\vec{G}$, $\vec{C}$ in vector form.

\medskip

\begin{figure*}[htb]
\centering

\tikzset{add/.style n args={4}{
    minimum width=6mm,
    path picture={
        \draw[circle] 
            (path picture bounding box.south east) -- (path picture bounding box.north west)
            (path picture bounding box.south west) -- (path picture bounding box.north east);
        \node[draw=none] at ($(path picture bounding box.south)+(0,0.13)$)     {\small #1};
        \node[draw=none] at ($(path picture bounding box.west)+(0.13,0)$)      {\small #2};
        \node[draw=none] at ($(path picture bounding box.north)+(0,-0.13)$)    {\small #3};
        \node[draw=none] at ($(path picture bounding box.east)+(-0.13,0)$)     {\small #4};
        }
    }
 }

\resizebox{1.0\textwidth}{!}{
 \begin{tikzpicture}[scale=0.6,-,>=stealth',shorten >=.2pt,auto,
     semithick, initial text=, ampersand replacement=\&,]

  \matrix[nodes={draw, fill=none, shape=rectangle, minimum height=.2cm, minimum width=.2cm, align=center}, row sep=.6cm, column sep=.6cm] {
    \node[draw=none] (r) {$R(z)$};
   \& \node[circle,add={-}{+}{}{}] (circle) {};
   \node[draw=none] (ez) at ([xshift=1cm,yshift=.15cm]circle)  {$e(z)$};
   \& \node[rectangle,draw,
	minimum width=1cm,
	minimum height=1cm,
        label=\textbf{Controller}] (cz) {\sc $\hat{C}(z)$};
   \node[draw=none] (ud) at ([xshift=1cm,yshift=.15cm]cz)  {$U(z)$};
     
   \& complexnode/.pic={ 
      \node[rectangle,draw,
	minimum width=3cm,
	minimum height=1.6cm,
	label=\textbf{DAC},] (dac) {};
     \node[circle,add={}{+}{+}{},fill=gray!20] (q2) at ([xshift=-.65cm]dac.center) {};
     \node[draw=none] (q2t)  at ([xshift=-.65cm,yshift=-.65cm]dac.center) {{\sc Q2}};
     \node[draw=none] (v2)  at ([xshift=-.65cm,yshift=1.5cm]dac.center) {$\nu_2(z)$};
     \node[fill=gray!20] (zoh) at ([xshift=.65cm]dac.center) {{\sc ZOH}};}   
   \& \node[rectangle,draw,
	minimum width=1cm,
	minimum height=1cm,
        label=\textbf{Plant}] (gs) {{\sc $\hat{G}(s)$}};
   \node[draw=none] (ud) at ([xshift=-2cm,yshift=.15cm]gs)  {$U(s)$};
   \node[draw=none] (y) at ([xshift=2cm,yshift=.15cm]gs)  {$\hat{Y}(s)$};
   \& complexnode/.pic={ 
     \node[rectangle,draw,
	minimum width=4cm,
	minimum height=1.6cm,
	label=\textbf{ADC},] (adc) {};
   \draw[] ([xshift=-1cm]adc.center) -- ++(0.5,0.2cm);
   \coordinate (switch1) at ([xshift=-1cm]adc.center);
   \coordinate (switch2) at ([xshift=-0.4cm]adc.center);
   \node[circle,add={}{+}{+}{},fill=gray!20] (q1) at ([xshift=1cm]adc.center) {};} 
     \node[draw=none] (q2t)  at ([xshift=1cm,yshift=-.65cm]adc.center) {{\sc Q1}};
   \node[draw=none] (v1)  at ([xshift=1cm,yshift=1.5cm]adc.center) {$\nu_1(z)$};
   \& \coordinate (aux1);
   \& \node[draw=none] (yz) {$\hat {Y}(z)$};\\
   \& \coordinate (aux3); 
   \&
   \&
   \& 
   \& 
   \& \coordinate (aux2);\\
  };

  \path[->] (v1) edge (q1.north);
  \path[->] (v2) edge (q2.north);
  \path[->] (r) edge (circle.west);
  \path[->] (aux1) edge (yz);
  \path  
   (circle.east) edge (cz)
   (cz.east) edge (q2.west)
   (q2.east) edge (zoh.west)
   (zoh.east) edge (gs.west)
   (switch2) edge (q1.west)
   (q1.east) edge (aux1.west)
   (gs.east) edge (switch1.west)
   (aux1.south) edge (aux2.north)
   (aux2.west) edge (aux3.east); 
  \path[->]  (aux3.north) edge (circle.south);
 \end{tikzpicture}
}
 \caption{Closed-loop digital control system (cf.~Section \ref{verifying-closed-loop-control-systems} for notation) \label{fig:sampledsystem}}
\end{figure*}
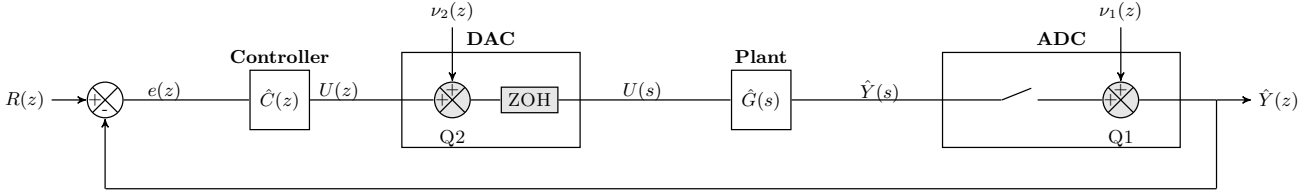

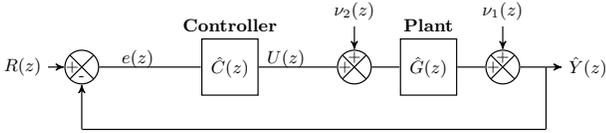
\begin{figure}[ht]
\centering
\tikzset{add/.style n args={4}{
    minimum width=6mm,
    path picture={
        \draw[circle] 
            (path picture bounding box.south east) -- (path picture bounding box.north west)
            (path picture bounding box.south west) -- (path picture bounding box.north east);
        \node[draw=none] at ($(path picture bounding box.south)+(0,0.13)$)     {\small #1};
        \node[draw=none] at ($(path picture bounding box.west)+(0.13,0)$)      {\small #2};
        \node[draw=none] at ($(path picture bounding box.north)+(0,-0.13)$)    {\small #3};
        \node[draw=none] at ($(path picture bounding box.east)+(-0.13,0)$)     {\small #4};
        }
    }
 }

\resizebox{.48\textwidth}{!}{
 \begin{tikzpicture}[scale=0.3,-,>=stealth',shorten >=.2pt,auto,
     semithick, initial text=, ampersand replacement=\&,]

  \matrix[nodes={draw, fill=none, shape=rectangle, minimum height=.2cm, minimum width=.2cm, align=center}, row sep=.6cm, column sep=.3cm] {
    \node[draw=none] (r) {$R(z)$};
   \& \node[circle,add={-}{+}{}{}] (circle) {};
   \node[draw=none] (ez) at ([xshift=1cm,yshift=.15cm]circle)  {$e(z)$};
   \& \node[rectangle,draw,
	minimum width=1cm,
	minimum height=1cm,
        label=\textbf{Controller}] (cz) {\sc $\hat{C}(z)$};
   \node[draw=none] (ud) at ([xshift=1cm,yshift=.15cm]cz)  {$U(z)$};
   \& \node[circle,add={}{+}{+}{}] (circle2) {};
   \node[draw=none] (nu2) at ([yshift=1cm]circle2)  {$\nu_2(z)$};
   \& \node[rectangle,draw,
	minimum width=1cm,
	minimum height=1cm,
        label=\textbf{Plant}] (gs) {{\sc $\hat{G}(z)$}};

   \& \node[circle,add={}{+}{+}{}] (circle3) {};
   \node[draw=none] (nu1) at ([yshift=1cm]circle3)  {$\nu_1(z)$};
   \& \coordinate (aux1);
   \& \node[draw=none] (yz) {$\hat Y(z)$};\\
   \& \coordinate (aux3); 
   \&
   \& 
   \& 
   \&
   \& \coordinate (aux2);\\
  };

  \path[->] (r) edge (circle.west);
  \path[->] (nu2) edge (circle2.north);
  \path[->] (nu1) edge (circle3.north);
  \path[->] (aux1) edge (yz);
  \path  
   (circle.east) edge (cz)
   (cz.east) edge (circle2.west)
   (circle2.east) edge (gs.west)
   (gs.east) edge (circle3.west)
   (circle3.east) edge (aux1.west)
   (aux1.south) edge (aux2.north)
   (aux2.west) edge (aux3.east); 
  \path[->]  (aux3.north) edge (circle.south);
 \end{tikzpicture}
}
\caption{Fully digital equivalent to system in Figure~\ref{fig:sampledsystem} 
\label{fig:digsystem1}}
\end{figure}

A typical digital control system with a continuous plant and a discrete
controller is illustrated in Figure~\ref{fig:sampledsystem}.  The DAC and
ADC converters introduce quantization errors (notice that each of them may
have a different FWL representation than the controller), which are modeled
as disturbances $\nu_{1}(z)$ and $\nu_{2}(z)$; $G(s)$ is the continuous-time
plant model with parametric additive uncertainty $\Delta_p{G}_n(s)$ and
$\Delta_p{G}_d(s)$ (as mentioned above); $R(z)$ is a given reference signal;
$U(z)$ is the control signal; and $\hat{Y}(z)$ is the output signal affected
by the disturbances and uncertainties in the closed-loop system
The ADC and DAC may be abstracted by transforming the closed-loop system in
Figure~\ref{fig:sampledsystem} into the digital system in 
Figure~\ref{fig:digsystem1}, 
where 
the effect of $\nu_{1}$ and $\nu_{2}$ in the output $Y(z)$ is
additive noise.  

In Figure~\ref{fig:digsystem1}, two sources of uncertainty are illustrated:
parametric uncertainties model the errors (which are represented
by~$\Delta_p \vec{G}$), and uncertainties for the quantizations in the ADC
and DAC conversions ($\nu_1$ and $\nu_2$), which are assumed to be
non-deterministic.  Recall that we discussed how the quantization noise is
an additive term, which means it does not enter parametrically in the
transfer function.  Instead, we later show that the system is stable given
these non-deterministic disturbance.

The uncertain model may be rewritten as a vector of coefficients in the
z-domain using equation \eqref{eq:hatgvector} as
$\vec{\hat{G}}=\vec{G}+\Delta_p \vec{G}$.  The parametric uncertainties in
the plant are assumed to have the same order as the plant model, since
errors of higher order can move the closed-loop poles by large amounts, thus
preventing any given controller from stabilizing such a setup.  This is a
reasonable assumption since most tolerances do not change the architecture
of the plant.

\smallskip

\paragraph{Direct use of controllers in fixed-point representation}

Since the controller is implemented using finite representation, $C(z)$ also
suffers disturbances from the FWL effects, with roundoffs in coefficients
that may change closed-loop poles and zeros position, and consequently
affect its stability, as argued in~\cite{Bessa16}.

Let $\hat{C}(z)$ be the digital controller transfer function represented
using this FWL with integer size $I$ and fractional size~$F$. 
The term $I$ affects the range of the representation and is set to avoid overflows, 
while $F$ affects the precision and the truncation after arithmetic operations. 
We shall denote the FWL domain of the coefficients by $\mathbb{R}\langle I,F
\rangle$ and define a function
\begin{align}
\small
\mathcal{F}_n&\param{I}{F}(P \in \mathcal{P}):\mathcal{P}^{n}\rightarrow \mathcal{P}^{n}\param{I}{F} \\
&\triangleq \tilde P \in \mathcal{P}^{n}\param{I}{F} : c_i \in \vec{P} \wedge \tilde c_i \in \vec{\tilde{P}}=\mathcal{F}_0\param{I}{F}(c_i),  \nonumber
\end{align}
where $\mathcal{P}^{n}$ is the space of polynomials of $n$-th order,
$\mathcal{P}^{n}\param{I}{F}$ is the space of polynomials with coefficients
in $\mathbb{R}\param{I}{F}$, and (as a special case)
$\mathcal{F}_0\param{I}{F}(x)$ returns the element $\tilde{x} \in
\mathbb{R}\param{I}{F}$ that is closest to the real parameter $x$.

Similarly, $\mathcal{F}_{n,m}\langle I,F \rangle(\cdot):\mathcal{P}^{n,m}\rightarrow \mathcal{P}^{n,m}\langle I,F \rangle$ applies the same effect to a ratio of polynomials, where $\mathcal{P}^{n,m}$,  $\mathcal{P}^{n,m}\langle I,F
\rangle$ are rational polynomial domains.

Thus, the perturbed controller model $\tilde{C}(z)$ may be obtained from the
original model $\hat{C}(z) = C(z) = \frac{C_{n}(z)}{C_{d}(z)}$ as follows:
\begin{equation}
\tilde{C}(z)=\mathcal{F}_{M_C,N_C}\param{I}{F}(C(z))=\frac{\mathcal{F}_{M_C}\param{I}{F} (\hat{C}_n(z))}{\mathcal{F}_{N_C}\param{I}{F}(\hat{C}_d(z))}.
\end{equation}
In the case of a digitally synthesized controller (as it is the case in this
work), $\tilde{C}(z) \equiv \hat{C}(z) \equiv C(z)$ because the synthesis is
performed directly using FWL representation.  In other words, we synthesize
a controller that is already in the domain $\mathbb{R} \param{I}{F}$ and has
therefore no uncertainties entering because of FWL representations, that is,
$\Delta_bC_n(z)=\Delta_bC_d(z)=0$.

\paragraph{Fixed-point computation in program synthesis}

The program synthesis engine uses fixed-point arithmetic.  Specifically, we
use the domain $\mathbb{R}\param{I}{F}$ for the controller's coefficients
and the domain $\mathbb{R}\param{I_p}{F_p}$ for the plant's coefficients,
where $I$ and $F$, as well as $I_p$ and $F_p$, denote the number of bits for
the integer and fractional parts, respectively, and where it is practically
motivated to consider $\mathbb{R}\param{ I_p}{F_p} \supseteq
\mathbb{R}\param{I}{F}$.

Given the use of fixed-point arithmetic, we examine the discretization effect 
during these operations. Let $\tilde C(z)$ and $\tilde G(z)$ be 
transfer functions represented using fixed-point bit-vectors.
\begin{align}
\small
\label{digital_controller_plant_tf}
\tilde C(z)&=\frac{\tilde \beta_{0}+\tilde \beta_{1}z^{-1}+...+\tilde \beta_{M_C}z^{-M_C}}{\tilde \alpha_{0}+\tilde \alpha_{1}z^{-1}+...+\tilde \alpha_{N_C}z^{-N_C}}, \\
\tilde G(z)&=\frac{\tilde b_{0}+\tilde b_{1}z^{-1}+...+\tilde b_{M_G}z^{-M_G}}{\tilde a_{0}+\tilde a_{1}z^{-1}+...+\tilde a_{N_G}z^{-N_G}}.
\end{align}

Recall that since the controller is synthesized in the $\mathbb{R}\param {I}{F}$
domain, $\tilde{C}(z) \equiv \hat{C}(z) \equiv C(z)$. 
However, given a real plant $\hat{G}(z)$, 
we need to introduce $\tilde G(z)=\mathcal{F}_{M_G,N_G}\param {I_p}{F_p}(\hat{G}(z))$, where 
\begin{align}
\label{digital_plant_tf}
\tilde G(z)&=\frac{(\hat{b}_{0}+\Delta_b \hat{b}_{0}) +...+(\hat{b}_{M_G}+\Delta_b \hat{b}_{M_G})z^{-M_G}}{(\hat{a}_{0}+\Delta_b \hat{a}_{0})+...+(\hat{a}_{N_G}+\Delta_b \hat{a}_{N_G})z^{-N_G}} \nonumber \\
\vec{\tilde G} &=\vec{\hat{G}}+\Delta_b{\vec{G}}=\vec{G}+\Delta_p{\vec{G}}+\Delta_b{\vec{G}}, 
\end{align}
where $\Delta_bc_i=\tilde{c}_i-\hat{c}_i$, and $\Delta_b{G}$ represents the
plant uncertainty caused by the rounding off effect.  We capture the global
uncertainty as $\Delta{\vec{G}}=\Delta_p{\vec{G}}+\Delta_b{\vec{G}}$.

\subsection{Closed-Loop Stability Verification under Parametric
Uncertainties, FWL Representation and Quantization Noise}
\label{sec:stability}

\aabatecmt{[Dario: please go through this section again, and clarify what sources of imprecision we consider where, distinguishing between plant and controller. also comment on why we operate on hat G rather than on tilde G - as we discussed yesterday. ]}
\dariocmt{I added a paragraph at the end of the section that should clarify.}
Sound synthesis of the digital controller requires the consideration of the
effect of FWL on the controller, and of quantization disturbances in the
closed-loop system.  Let the quantizer $Q1$ (ADC) be the source of a white
noise $\nu_{1}$, and $Q2$ (DAC) be the source of a white noise~$\nu_{2}$. 
The following equation models the system in Figure~\ref{fig:sampledsystem},
including the parametric uncertainties $\Delta \vec{G}$ and the FWL
effects on the controller $\tilde{C}(z)$:
\begin{equation}
\hat{Y}(z)=\nu_{1}(z)+\hat{G}(z)\tilde{C}(z)R(z)+\hat{G}(z)\nu_{2}(z)-\hat{G}(z)\tilde{C}(z)\hat{Y}(z). 
\end{equation}
The above can be rewritten as follows:
\begin{equation}
\label{eq:outputfunctions}
\hat{Y}(z)=H_{1}(z)\nu_{1}(z)+H_{2}(z)\nu_{2}(z)+H_{3}(z) R(z),
\end{equation}
where
\begin{equation*}
H_{1}(z)=\frac{1}{1+\hat{G}(z) \tilde{C}(z)}, 
\end{equation*}
\begin{equation*}
H_{2}(z)=\frac{\hat{G}(z)}{1+\hat{G}(z) \tilde{C}(z)}, 
\end{equation*}
\begin{equation*}
H_{3}(z)=\frac{\hat{G}(z) \tilde{C}(z)}{1+\hat{G}(z) \tilde{C}(z)}. 
\end{equation*}

\begin{myassumption}
\label{whitenoise}
The quantization noises $\nu_{1}$ (from $Q1$) and $\nu_{2}$ (from $Q2$) are
uncorrelated white noises and their amplitudes are always bounded by the
half of quantization step~\cite{astrom1997computer}, i.e., $\vert \nu_{1}
\vert \leq \frac{q_{1}}{2}$ and $\vert \nu_{2} \vert \leq \frac{q_{2}}{2}$,
where $q_{1}$ and $q_{2}$ are the quantization steps of ADC and DAC, respectively.
\end{myassumption}

A discrete-time dynamical system is said to be Bounded-Input and
Bounded-Output (BIBO) stable if bounded inputs necessarily result in bounded
outputs.  This condition holds true over an LTI model if and only if every
pole of its transfer function lies inside the unit circle \cite{Astrom08}. 
Analyzing Eq.~\eqref{eq:outputfunctions}, the following proposition provides
conditions for the BIBO stability of the system in
Figure~\ref{fig:sampledsystem}, with regards to the exogenous signals
$R(z)$, $\nu_{1}$, and $\nu_{2}$, which are all bounded (in particular, the
bound on the quantization noise is given by Assumption~\ref{whitenoise}).
\begin{myprop}{\cite{Bessa16,fadali}}
\label{prop:eq_int_stab} 
Consider a feedback closed-loop control system as given in
Figure~\ref{fig:sampledsystem} with a FWL implementation of the digital
controller $\tilde{C}(z) =\mathcal{F}_{M_C,N_C}\param {I}{F}(C(z))$ and
uncertain discrete model of the plant
from~\eqref{eq:coefficients}, \eqref{eq:delta_coefficients}
$$
\hat{G}(z) =\frac{\hat{G}_n(z)}{\hat{G}_d(z)}, \quad \vec{\hat{G}}=\vec{G}+\Delta_p \vec{G}.
$$
Then $\hat{G}(z)$ is BIBO-stable if and only if:
\begin{itemize*}
\item  the roots of characteristic polynomial $S(z)$ are inside the open unit circle, where $S(z)$ is:
\begin{equation}
\label{eq:internal_stab_lemma}
\hspace{-3ex} S(z)=\tilde{C}_n(z)\hat{G}_n(z)+\tilde{C}_d(z)\hat{G}_d(z);
\end{equation}
\item the direct loop product $\tilde{C}(z) \hat{G}(z)$ has no pole-zero cancellation on or outside the unit circle.
\end{itemize*}
\end{myprop}

Proposition~\ref{prop:eq_int_stab} provides necessary (and sufficient)
conditions for the controller to stabilize the closed-loop system,
considering plant parametric uncertainties (i.e., $\Delta_p \vec{G}$),
quantization noises ($\nu_{1}$ and $\nu_{2}$) and FWL effects in the control
software.  In particular, note that the model for quantization noise enters
as a signal to be stabilized: in practice, if the quantization noise is
bounded, the noise may be disregarded if the conditions on
Proposition~\ref{prop:eq_int_stab} are satisfied.

If the verification is performed using FWL arithmetic,  the above equations
must use $\tilde{G}(z)$ instead of $\hat{G}(z)$.  The former will provide
sufficient conditions for the latter to be stabilized.

\section{Automated Program Synthesis \\ for Digital
Control}\label{sec:synthesis}

\subsection{Overview of the Synthesis Process}
\label{verification-flow}

\begin{figure*}[t]
\centering
\includegraphics[width=0.9\textwidth]{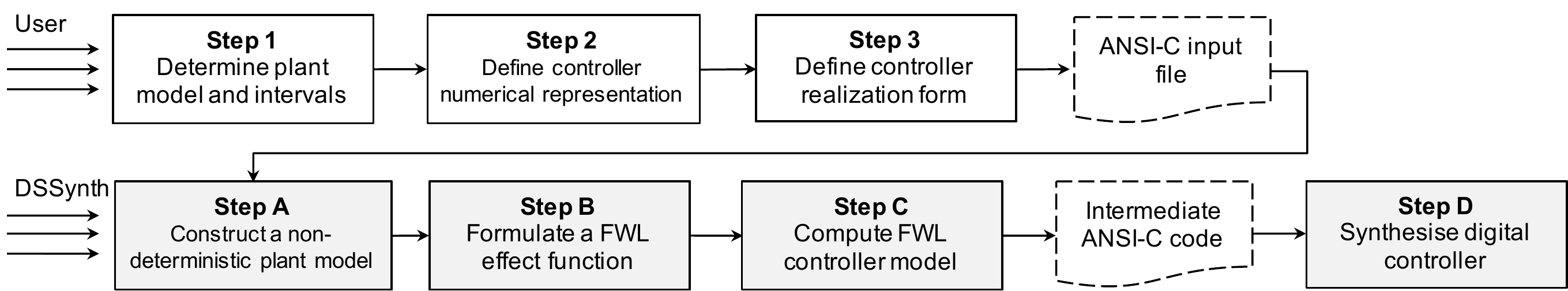}
\vspace{0.1cm}
\caption{Overview of the synthesis process\label{DSSynth_process}}
\end{figure*}

In order to synthesize closed-loop digital control systems, we use a program
synthesis engine.  Our program synthesizer implements Counter-Example Guided
Inductive Synthesis (CEGIS)~\cite{sketch}.  We start by presenting its
general architecture followed by describing the parts specific to
closed-loop control systems.  A high-level view of the synthesis process is
given in Figure~\ref{DSSynth_process}.  Steps $1$ to $3$ are performed by
the user and Steps~A to D are automatically performed by our tool for
Digital Systems Synthesis, named \tool.

CEGIS-based control synthesis requires a formal verifier to check whether a
candidate controller meets the requirements when combined with the plant. 
We use the Digital-System Verifier (DSVerifier)~\cite{IsmailBCFF15} in the
verification module for \tool.  It checks the stability of closed-loop control systems and
considers finite-word length (FWL) effects in the digital controller, and
uncertainty parameters in the plant model (plant intervals)~\cite{Bessa16}.   

%

Given a plant model in ANSI-C syntax as input (Steps 1--3), \tool constructs
a non-deterministic model to represent the plant family, i.e., it addresses
plant variations as interval sets (Step~A), and formulates a function
(Step~B) using implementation details provided in Steps~$2$ and $3$ to
calculate the controller parameters to be synthesized (Step~C).  Note that
\tool synthesizes the controller for the desired numerical representation
and realization form.  Finally, \tool builds an intermediate ANSI-C code for
the digital system implementation, which is used as input for the CEGIS
engine (Step~D).

This intermediate ANSI-C code model contains a specification $\phi$ for the
property of interest (i.e., robust stability) and is passed to the
Counterexample-Guided Inductive Synthesis (CEGIS) module of
CBMC~\cite{ClarkeKL04}, where the controller is marked as the input variable
to synthesize.  CEGIS employs an iterative, counterexample-guided refinement
process, which is explained in detail in Section~\ref{synthesizer-general}. 
CEGIS reports a successful synthesis result if it generates a controller
that is safe with respect to~$\phi$.  In particular, the ANSI-C code model
guarantees that a synthesized solution is complete and sound with respect to
the stability property~$\phi$, since it does not depend on system inputs and
outputs.  In the case of stability, the specification~$\phi$ consists of a
number of assumptions on the polynomial coefficients, following Jury's
Criteria, as well as the restrictions on the representation of these
coefficients as discussed in detail in Section~\ref{synthesis-elements}.

\subsection{Architecture of the Program Synthesizer}
\label{synthesizer-general}
%
%
The input specification provided to the program synthesizer is of the form
$\exists \vec{P} .  \forall \vec{x}.  \sigma(\vec{x}, \vec{P})$ where
$\vec{P}$ ranges over functions, $\vec{x}$ ranges over ground terms and
$\sigma$ is a quantifier-free formula.  We interpret the ground terms over
some finite domain $\mathcal{D}$.

The design of our synthesizer is given in Figure~\ref{fig:CEGIS} and consists
of two phases, {\sc Synthesize} and {\sc Verify}, which interact via a
finite set of test vectors {\sc inputs} that is updated incrementally. 
Given the aforementioned specification $\sigma$, the {\sc synth} procedure
tries to find an existential witness $\vec{P}$ satisfying the specification
$\sigma(\vec{x}, \vec{P})$ for all $\vec{x}$ in {\sc inputs} (as opposed to
all $\vec{x} \in \mathcal{D}$).
If {\sc synthesize} succeeds in finding a witness~$\vec{P}$, this witness
is a candidate solution to the full synthesis formula.  We pass this
candidate solution to {\sc verify}, which checks whether it is a full
solution (i.e., $\vec{P}$ satisfies the specification $\sigma(\vec{x},
\vec{P})$ for all $\vec{x}\in\mathcal{D}$).
%
%
%
If this is the case, then the algorithm terminates.  Otherwise, additional
information is provided to the {\sc synthesize} phase in the form of a new
counterexample that is added to the {\sc inputs} set and the loop iterates
again (the second feedback signal ``Increase Precision'' provided by the
{\sc Verify} phase in Figure~\ref{fig:CEGIS} is specific to control
synthesis and will be described in the next section).


Each iteration of the loop adds a new input to the finite set $\text{\sc
inputs}$ that is used for synthesis.  Given that the full set of inputs
$\mathcal{D}$ is finite, this means that the refinement loop can only
iterate a potentially very large, but finite number of times.

\subsection{Synthesis for Control}
\label{synthesis-elements}

\paragraph{Formal specification of the stability property}

Next, we describe the specific property that we pass to the program
synthesizer as the specification $\sigma$.  There are a number of 
algorithms in our verification engine that can be used for stability analysis~\cite{daes20161, Bessa16}.  
Here we choose Jury's criterion~\cite{astrom1997computer} in view of its efficiency 
\aabatecmt{[check correctness]} 
and ease of integration within \tool: 
we employ this method to check the stability in the $z$-domain for the
characteristic polynomial $S(z)$ defined in~\eqref{eq:internal_stab_lemma}.
%
%
We~consider the following form for $S(z)$:
\begin{equation*}
S(z) = a_0z^N+a_1z^{N-1}+\cdots+a_{N-1}z+a_N=0, a_0\neq0. 
\end{equation*}

Next, the following matrix
$M = [m_{ij}]_{(2N-2)\times N}$ is built from $S(z)$ coefficients:
$$
M=\left( 
\begin{array}{c}
V^{(0)}\\
V^{(1)}\\
\vdots\\
V^{(N-2)}
\end{array}
\right), 
$$
where $V^{(k)} = [v^{(k)}_{ij} ]_{2\times N}$ such that:
$$
v_{ij}^{(0)}=\left\{
\begin{array}{ll}
a_{j-1}, & \texttt{if}~i=1\\
v_{(1)(N-j+1)}^{0},&\texttt{if}~i=2
\end{array}
\right.
$$
$$
v_{ij}^{(k)}=\left\{
\begin{array}{ll}
0,&\texttt{if}~j>n-k\\
v_{1j}^{(k-1)}-v_{2j}^{(k-1)} . \frac{v_{11}^{(k-1)}}{v_{21}^{(k-1)}}, & \texttt{if}~j\leq n-k ~\texttt{and}~i=1\\
v_{(1)(N-j+1)}^{k},& \texttt{if}~j\leq n-k ~\texttt{and}~i=2\\
\end{array}
\right.
$$
and where $k \in \mathbb{Z}$ is such that $0 < k < N - 2$. 
We have that~\cite{astrom1997computer} 
$S(z)$ is the
characteristic polynomial of a stable system if and only if the following four conditions hold:
$R_1: S(1) > 0$;
$R_2: (−1)^N S(−1) > 0$;
$R_3: |a_0| < a_N$;
$R_4: m_{11} > 0 \wedge\allowbreak
      m_{31}>0 \wedge\allowbreak
      m_{51}>0 \wedge \ldots \wedge\allowbreak
      m_{(2N{-}3)(1)}>0$.
%
The stability property is then encoded by a constraint of the form:
$
\phi_\mathit{stability} \equiv (R_1 \wedge R_2 \wedge R_3 \wedge R_4).
$


\paragraph{The synthesis problem}
The synthesis problem we are trying to solve is the following:
find a digital controller $\tilde C(z)$ 
that makes the closed-loop system stable 
for all possible uncertainties 
$\tilde G(z)$ \eqref{digital_plant_tf}.
When mapping back to the notation used for describing the general architecture 
of the program synthesizer, the controller $\tilde C(z)$ denotes $P$ and 
$\tilde G(z)$ represents $x$. 

As mentioned above, we compute the coefficients for $\tilde C(z)$ 
in the domain $\mathbb{R}\param{I}{F}$, 
and those for $\tilde G(z)$ in the domain
$\mathbb{R}\langle I_p,F_p \rangle$.
While the controller's precision $\param{I}{F}$ is given, 
we can vary $\param{I_p}{F_p}$ such that 
$\mathbb{R}\langle I_p,F_p \rangle \supseteq \mathbb{R}\langle I,F \rangle$.
As~the cost of SAT solving increases with in the size of the problem
instance, our algorithm tries to solve the problem first for small $I_p$,
$F_p$, iteratively increasing the precision if it is insufficient.

\subsection{The {\sc Synthesize} and {\sc Verify} phases}

The {\sc synthesize} phase uses BMC to compute a solution $\tilde C(z)$. 
There are two alternatives for the {\sc verify} phase.  The first approach
uses interval arithmetic \cite{moore1966interval} to represent the
coefficients $[{c}_i-\Delta_p{c}_i-\Delta_b{c}_i,\,\allowbreak
{c}_i+\Delta_p{c}_i-\Delta_b{c}_i+(2^{-F_p})]$ and rounds outwards.  This0
allows us to simultaneously evaluate the full collection of plants
$\hat{G}(s)$ (i.e., all concrete plants $G(s)$ in the range $G(s) \pm
\Delta_pG(s)$) plus the effects of numeric calculations.  Synthesized
controllers are stable for all plants in the family.  Preliminary
experiments show that a synthesis approach using this verification engine
has poor performance and we therefore designed a second approach.  Our
experimental results in Section~\ref{sec:experiments} show that the speedup
yielded by the second approach is in most cases of at least two orders of
magnitude.

\begin{figure*}[htb]
\centering
{
 \begin{tikzpicture}[scale=0.3,->,>=stealth',shorten >=.2pt,auto, semithick, initial text=, ampersand replacement=\&,]
  \matrix[nodes={draw, fill=none, shape=rectangle, minimum height=.2cm, minimum width=.2cm, align=center
},
          row sep=.6cm, column sep=2cm] {
   \coordinate (aux1);
   \& \coordinate (aux2);
   \&;\\
   \node[minimum width=2.5cm, minimum height=0.6cm, fill=gray!20] (synth) {{\sc Synthesize}};
   \&
   complexnode/.pic={ 
     \node[rectangle,draw,dotted,
	minimum width=7cm,
	minimum height=1cm,
        pattern=north west lines, pattern color=gray!20,
	label={\sc Verify},] (verif) {};
     \node[minimum width=2.5cm, minimum height=0.6cm, fill=gray!20] (verif1) at ([xshift=-2cm]verif.center) {{\sc Uncertainty}};
     \node[minimum width=2.5cm, minimum height=0.6cm, fill=gray!20] (verif2) at ([xshift=2cm]verif.center) {{\sc Precision}};
   } 
   \& \node[ellipse, fill=gray!20] (done) {{\sc Done}};\\
   \& \\
   \node[minimum height=0.5cm] (gp) {\sf Closed-loop System Search};
   \&
   complexnode/.pic={ 
     \coordinate (aux);
   \node[minimum height=0.5cm] (bmc) at ([xshift=-2cm]aux.center) {\sf BMC-based Verifier};
   \node[minimum height=0.5cm] (fp)  at ([xshift=2cm]aux.center) {\sf Fixed-point Arithmetic Verifier};
   }   
    \\
  };

   \path
    ([yshift=2em]synth.east) edge node[xshift=-0.5em] {Candidate solution} ([yshift=2em]verif1.west)
    ([yshift=-2em]verif1.west) edge node {Counterexample} ([yshift=-2em]synth.east)
    ([xshift=5em]verif1.south) edge node[align=center] {Candidate\\ $P$} ([xshift=5em]bmc.north)
    ([xshift=5em]verif2.south) edge node[align=center] {Candidate\\ $P$} ([xshift=5em]fp.north)
    ([xshift=-5em]bmc.north) edge node[align=center]  {UNSAT/\\model} ([xshift=-5em]verif1.south)
    ([xshift=-5em]fp.north) edge node[align=center]  {True/\\False} ([xshift=-5em]verif2.south)
    (verif2) edge node {} (done)
    ([xshift=5em]synth.south) edge node[align=center] {Inputs} ([xshift=5em]gp.north)
    ([xshift=-5em]gp.north) edge node[align=center] {UNSAT/\\candidate} ([xshift=-5em]synth.south)
    (aux1) edge (synth.north);
   \path[-]
   (verif2.north) edge node[align=center] {} ([xshift=6.7cm]aux2)
   ([xshift=6.7cm]aux2) edge node[align=center] {Increase Precision} (aux1);

 \end{tikzpicture}
}
\caption{Counterexample-Guided Inductive Synthesis of Closed-loop Systems (Step~D)
\label{fig:CEGIS}}
\end{figure*}
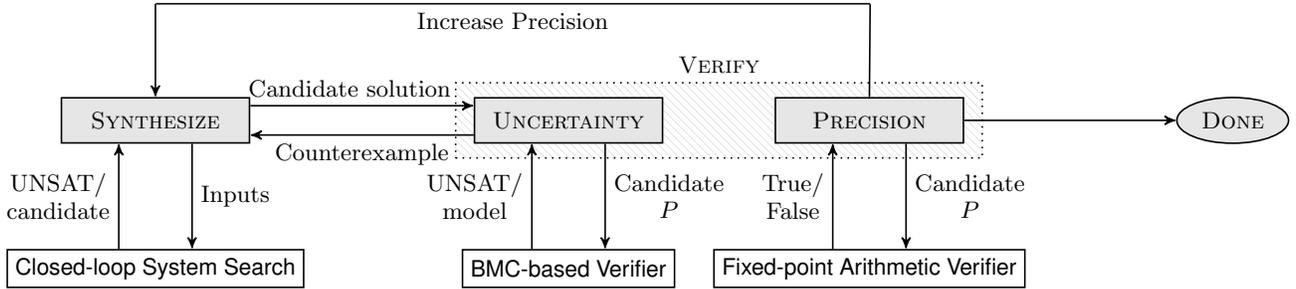

The second approach is illustrated in Figure~\ref{fig:CEGIS} and uses a
two-stage verification approach: the first stage performs potentially
unsound fixed-point operations assuming a plant precision \param{I_p}{F_p},
and the second stage restores soundness by validating these operations using
interval arithmetic on the synthesized controller.
In more detail, in the first stage, denoted by {\sc Uncertainty} in
Figure~\ref{fig:CEGIS}, assuming a precision \param{I_p}{F_p} we check
whether the system is unstable for the current candidate solution, i.e., if
$\neg \phi_\mathit{stability}$ is satisfiable for~$S(z)$.  If this is the
case, then we obtain a counterexample~$\tilde G(z)$,
%
%
which makes the closed-loop system unstable.  This uncertainty is added to
the set {\sc inputs} such that, in the subsequent {\sc synthesize} phase, we
obtain a candidate solution consisting of a controller $C(z)$, which makes
the closed-loop system stable for all the uncertainties accumulated in {\sc
inputs}.

If the {\sc Uncertainty} verification stage concludes that the system is
stable for the current candidate solution, then we pass this solution to the
second verification stage, {\sc Precision}, which checks the propagation of
the error in the fixed-point calculations using a Fixed-point Arithmetic
Verifier based on interval arithmetic.
%
%

If the {\sc precision} verification returns $\mathit{false}$, then we
increase the precision of \param{I_p}{F_p} and re-start the {\sc
  synthesize} phase with an empty {\sc inputs} set.  Otherwise, we
found a full sound solution for our synthesis problem and we are done.

In the rest of the paper, we will refer to the two approaches 
for the {\sc verify} phase as one-stage and two-stage, respectively.

\subsection{Soundness}

The {\sc synthesise} phase generates potentially unsound candidate
solutions.  The soundness of the model is ensured by the {\sc verify} phase. 
If a candidate solution passes verification, it is necessarily sound.

The {\sc verify} phase has two stages.  The first stage ensures that no
counterexample plant with an unstable closed loop exists over
finite-precision arithmetic.  Since the actual plant uses reals, we need to
ensure we do not miss a counterexample because of rounding errors.  For this
reason, the second verification stage uses an overapproximation with
interval arithmetic with outward rounding.  Thus, the first verification
stage underapproximates and is used to generate counterexamples, and the
second stage overapproximates and provides proof that no counterexample
exists.

\subsection{Illustrative Example} \label{sec:running-ex}

We illustrate our approach with a classical cruise control example from the
literature~\cite{Astrom08}.  It highlights the challenges that
arise when using finite-precision arithmetic in digital control.  We are
given a discrete plant model (with a time step of $0.2$\,s), 
represented by the following $z$-expression:
\begin{equation}
\label{Eq:running-example-plant}
G(z) = \frac{0.0264}{z-0.9998}.
\end{equation}

Using an optimization tool, the authors
of~\cite{DBLP:conf/hybrid/WangGRJF16} have designed a high-performance
controller for this plant, which is characterized by the following
$z$-domain transfer function:
\begin{equation}
\label{Eq:running-example-controller}
C(z) = \frac{2.72z^2 - 4.153z + 1.896}{z^2 - 1.844z + 0.8496}.
\end{equation}
The authors of~\cite{DBLP:conf/hybrid/WangGRJF16} claim that the controller
$C(z)$ in~\eqref{Eq:running-example-controller} stabilizes the closed-loop
system for the discrete plant model $G(z)$
in~\eqref{Eq:running-example-plant}.  However, if the effects of
finite-precision arithmetic are considered, then this closed-loop system
becomes unstable.
For instance, an implementation of $C(z)$ using 
$\mathbb R \param {4}{16}$ fixed-point
numbers (i.e., $4$ bits for the integer part and $16$ bits for the
fractional part) can be modeled as: 
\begin{equation}
\label{Eq:running-example-controller-quantized}
\resizebox{.47\textwidth}{!}{
$\tilde{C}(z) {:=} \frac{2.7199859619140625z^2{-}4.1529998779296875z
{+}1.89599609375}{z^2{-}1.843994140625z+0.8495941162109375}$. 
}
\end{equation} 
The resulting system, 
where $\tilde{C}(z)$ and $G(z)$ are in the forward path, is unstable. 
Notice that this is disregarding further approximation effects on the plant caused by quantization in the verifier (i.e., $\tilde{G}(z)$).
Figure~\ref{fig:original} gives the Bode diagram for the digital controller
represented in~\eqref{Eq:running-example-controller}: 
as the phase margin is negative, 
the controller is unstable when considering the FWL effects.

\subsection{Program Synthesis for the Example}
We now demonstrate how our approach solves the synthesis problem for the
example given in the previous section.  Assuming a precision of $I_p=16$,
$F_p=24$, we start with an a-priori candidate solution with all coefficients
zero (the controller performs FWL arithmetic, hence we use
$\tilde{C}(z)$):
$$ 
\tilde{C}(z)=\frac{0z^2{+}0z{+}0}{0z^2{+}0z{+}0}. 
$$
In the first {\sc verify} stage, the {\sc uncertainty} 
check finds the following counterexample:
$$ 
\tilde G(z) = \frac{0.026506}{1.000610z+1.002838}. 
$$
We add this counterexample to {\sc inputs} and initiate the {\sc synthesize}
phase, where we obtain the following candidate solution:
%
$$
\tilde{C}(z)=\frac{12.402664z^2{-}11.439667z{+}0.596756}{4.003906z^2{-}0.287949z{+}0.015625}. 
$$ 
This time, the {\sc uncertainty} check does not find any
counterexample and we pass the current candidate solution to the {\sc
precision} verification stage.
%
%
We obtain the result $\mathit{false}$, meaning that the current precision is
insufficient.  Consequently, we increase our precision to $I_p=20$, $F_p=28$.
Since the previous counterexamples were obtained at lower precision, we
remove them from the set of counterexamples.  Back in the {\sc synthesize}
phase, we re-start the process with a candidate solution with all
coefficients $0$, as above.  Next, the {\sc uncertainty} verification stage provides
the first counterexample at higher precision:
%
$$ 
\tilde G(z) = \frac{0.026314}{0.999024z{-}1.004785}. 
$$
In the {\sc synthesize} phase, we get a new candidate solution that
eliminates the new, higher precision counterexample:
$$ 
\tilde{C}(z)=\frac{11.035202z^2{+}5.846100z{+}4.901855}{1.097901z^2{+}0.063110z{+}0.128357}. 
$$
%

\begin{figure}[tb]
    \centering
    \begin{subfigure}[b]{0.45\textwidth}
        \includegraphics[width=\textwidth]{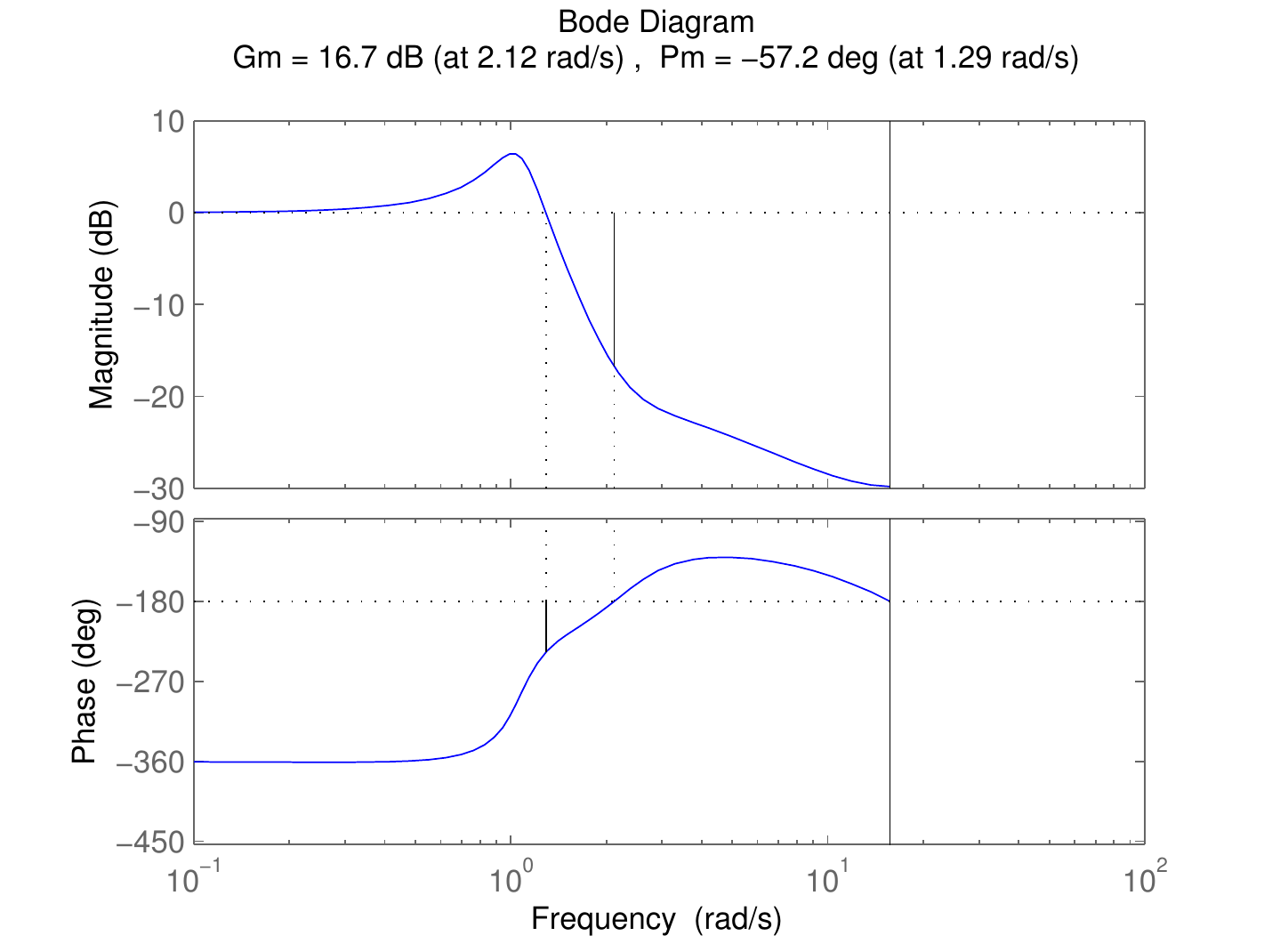}
        \caption{Original controller~\cite{DBLP:conf/hybrid/WangGRJF16}}
        \label{fig:original}
    \end{subfigure}
    ~
    \begin{subfigure}[b]{0.45\textwidth}
        \includegraphics[width=\textwidth]{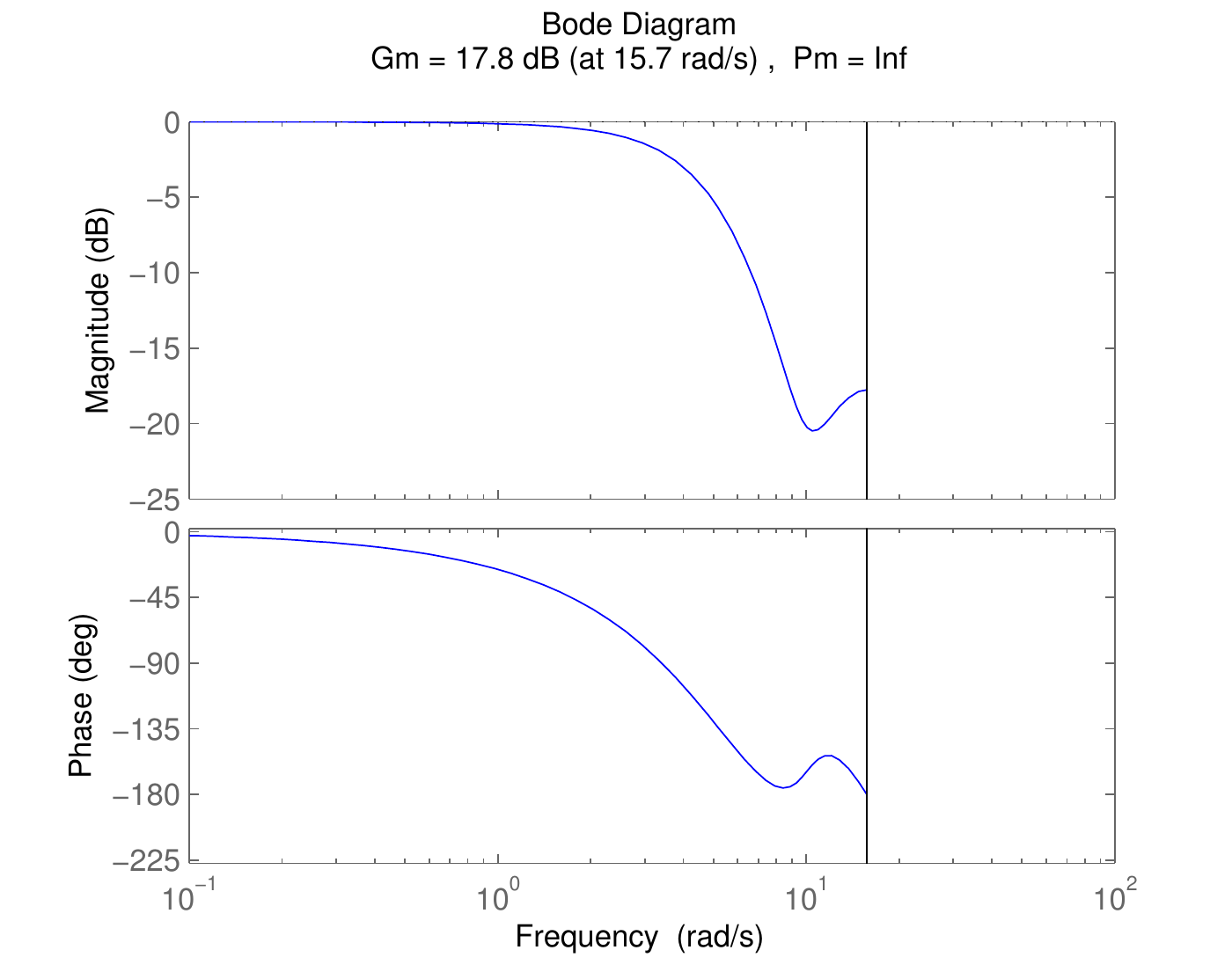}
        \caption{Controller synthesized by \tool}
        \label{fig:cegiscontroller}
    \end{subfigure}
    \caption{Bode diagram for original controller in~\cite{DBLP:conf/hybrid/WangGRJF16} 
    and for newly synthesized closed-loop system}\label{fig:bode}
\end{figure}

This candidate solution is validated  as the final solution by both stages
{\sc uncertainty} and {\sc precision} in the {\sc verify} phase. 
Figure~\ref{fig:bode} compares the Bode diagram using the digital controller 
represented by Eq.~\eqref{Eq:running-example-controller}
from~\cite{DBLP:conf/hybrid/WangGRJF16} (Figure~\ref{fig:original}) and the
final candidate solution from our synthesizer
(Figure~\ref{fig:cegiscontroller}).  The \tool final solution is stable
since it presents an infinite phase margin and a gain margin of $17.8$\,dB.





%

\begin{figure*}
    \centering
    \begin{subfigure}[b]{0.3\textwidth}
        \includegraphics[width=\textwidth]{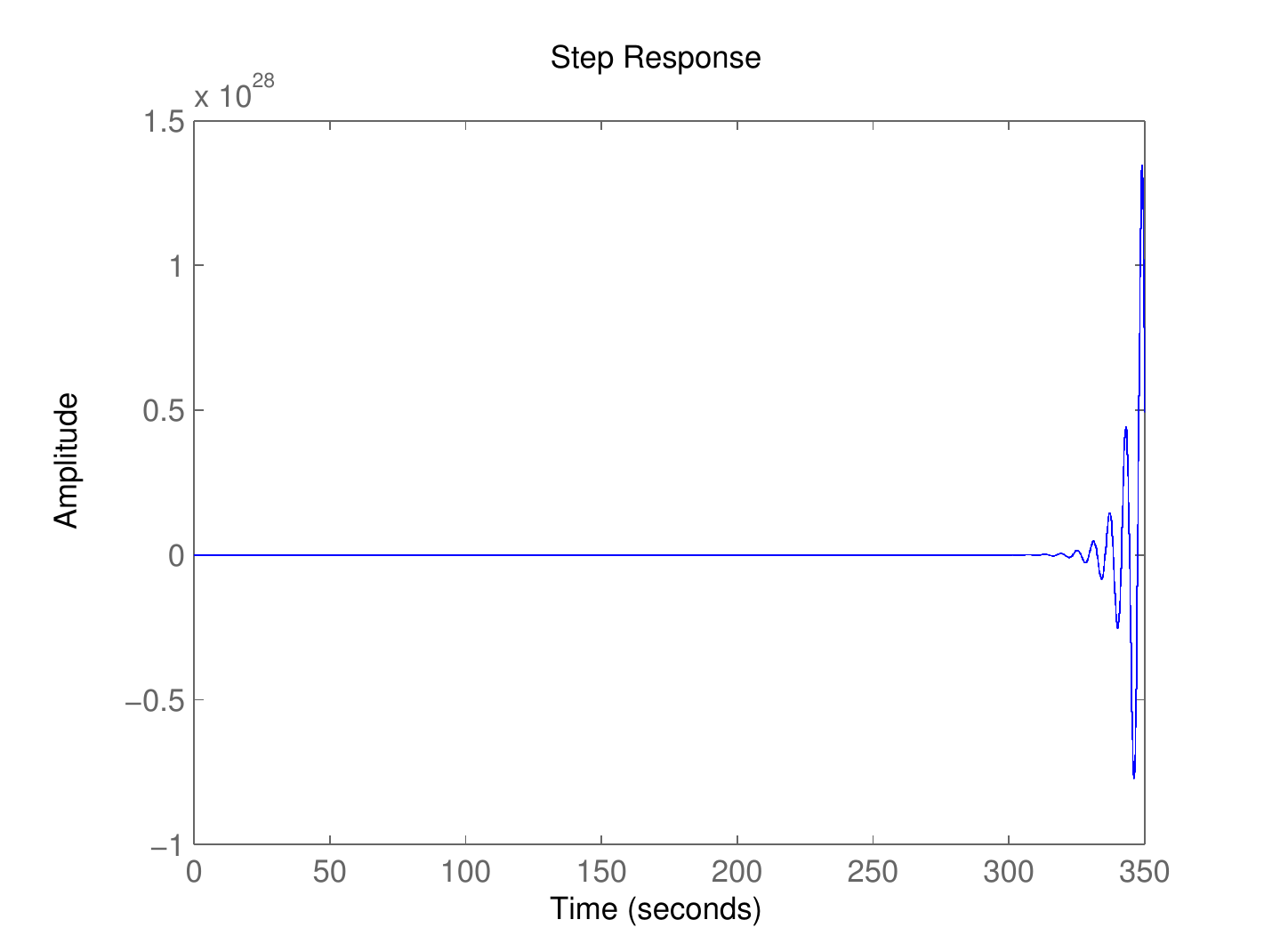}
        \caption{Original controller}
        \label{fig:step0}
    \end{subfigure}
    ~
    \begin{subfigure}[b]{0.3\textwidth}
        \includegraphics[width=\textwidth]{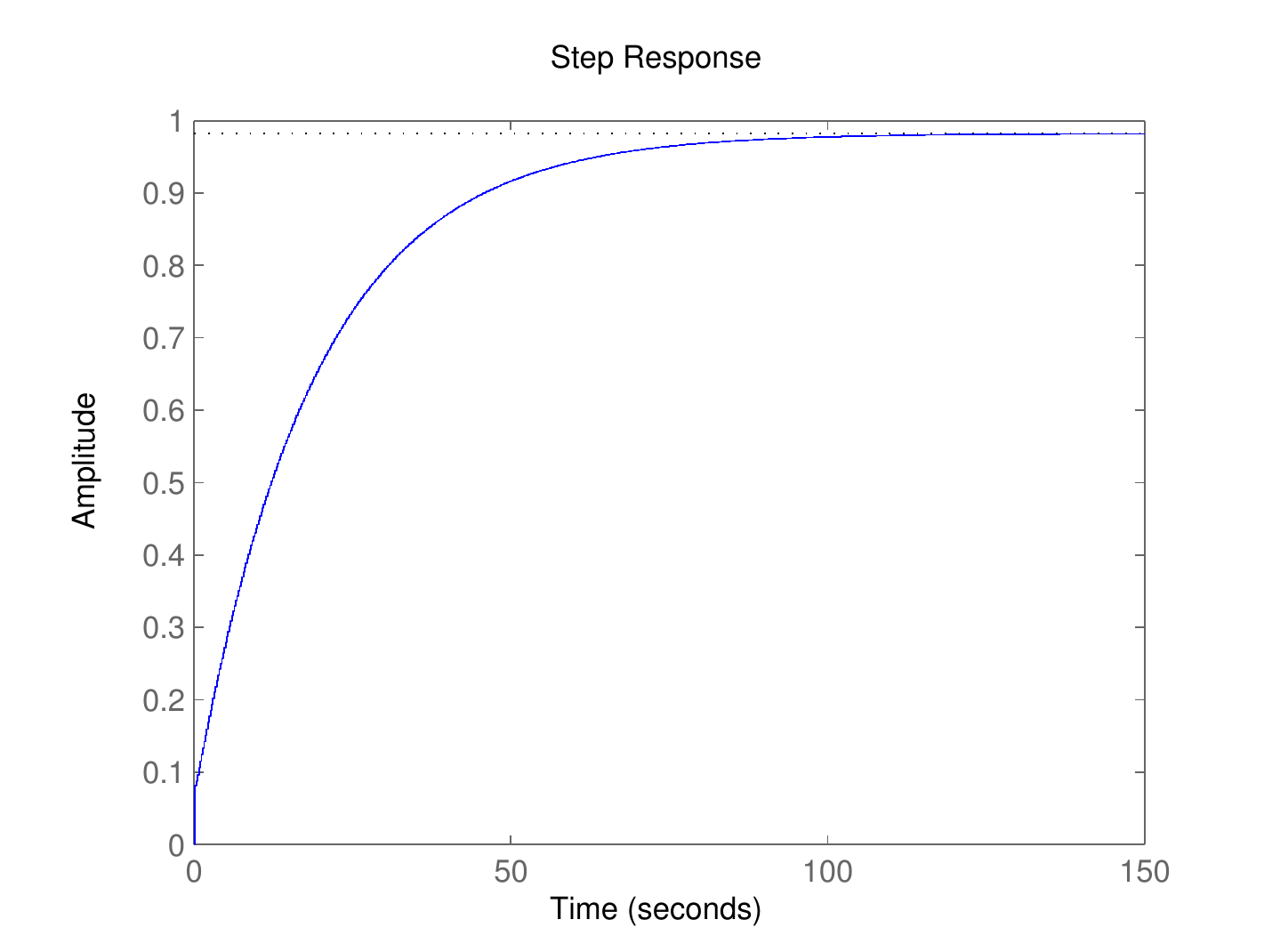}
        \caption{First solution by \tool}
        \label{fig:step1}
    \end{subfigure}
    ~
    \begin{subfigure}[b]{0.3\textwidth}
        \includegraphics[width=\textwidth]{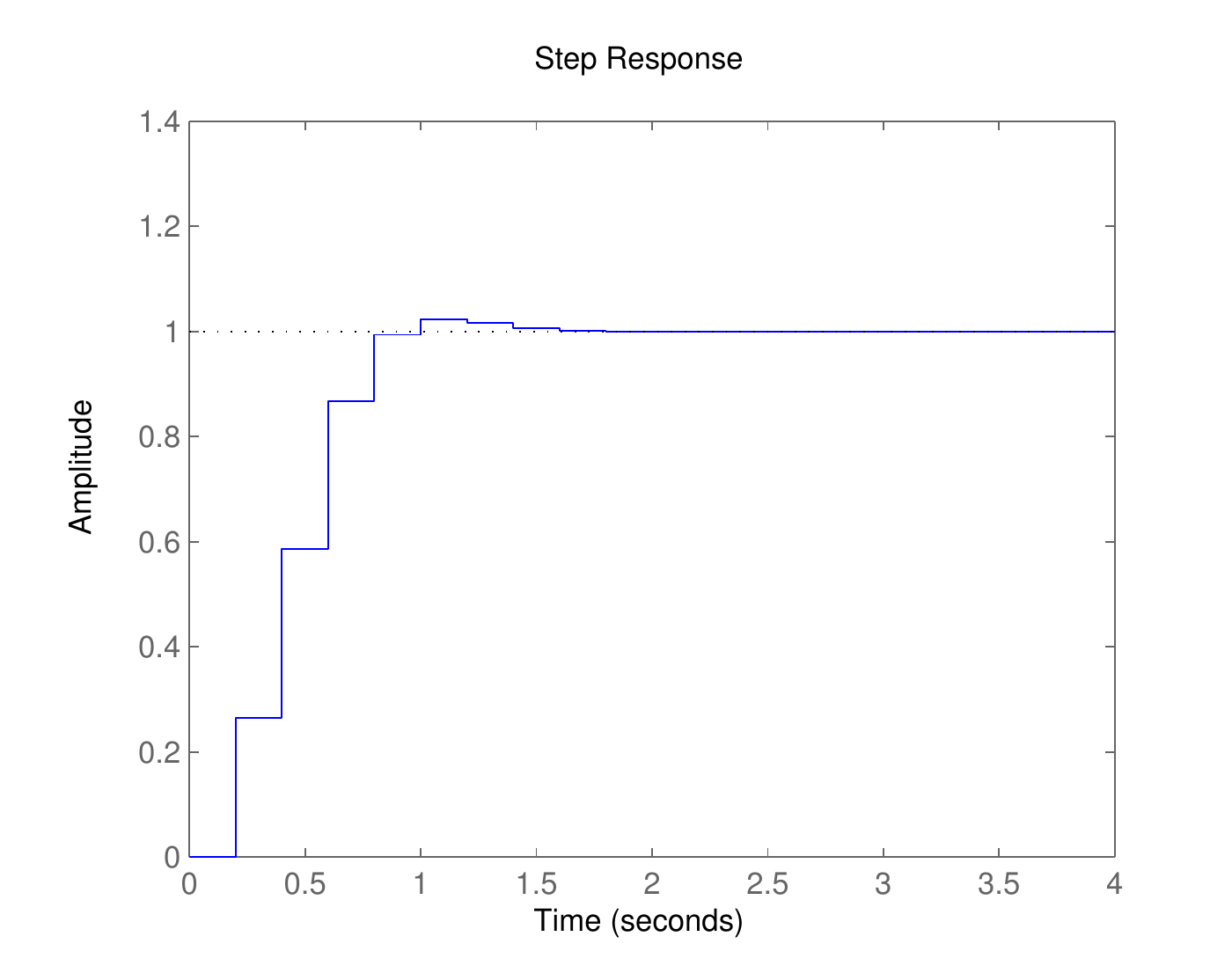}
        \caption{Final solution by \tool}
        \label{fig:step2}
    \end{subfigure}
    \caption{Step responses for original~\cite{DBLP:conf/hybrid/WangGRJF16}
             closed-loop system with FWL effects and for each
             {\sc synthesize} iteration of \tool}\label{fig:step}
\end{figure*}

Figure~\ref{fig:step} illustrates the step responses of the closed-loop
system with the original controller represented by
Eq.~\eqref{Eq:running-example-controller} (Figure~\ref{fig:step0}), the
first (Figure~\ref{fig:step1}) and final (Figure~\ref{fig:step2}) candidate
solutions provided by \tool.  The step response in Figure~\ref{fig:step0}
confirms the stability loss if we consider FWL effects.  
Figure~\ref{fig:step1} shows that the first candidate controller is able to
stabilize the closed-loop system without uncertainties, but it is rejected
during the {\sc precision} phase by \tool since this solution is not
sound.  Finally, Figure~\ref{fig:step2} shows a stable
behavior for the final (sound) solution, which presents a lower settling
time (hence the digitization effects).

\section{Experimental Evaluation}\label{sec:experiments}


\subsection{Description of the Benchmarks}
\label{experimental-setup}

The first set of benchmarks uses the discrete model $G_1$ of a cruise control
system for a car, and accounts for rolling friction, aerodynamic drag, and
the gravitational disturbance force~\cite{Astrom08}. 
%
%
The second set of benchmarks considers the discrete model $G_2$ 
of a simple spring-mass damper plant~\cite{DBLP:conf/hybrid/WangGRJF16}. 
%
%
A third set of benchmarks uses the discrete model $G_{3}$ for satellite attitude 
dynamics~\cite{Franklin15}, which require attitude control
for orientation of antennas and sensors w.r.t.~Earth.
The fourth set of benchmarks presents an alternative discrete model $G_4$ 
of a cruise control system~\cite{DBLP:conf/hybrid/WangGRJF16}. 
The fifth and sixth set of benchmarks describe the discrete model 
of a DC servo motor velocity dynamics~\cite{exampleCAD,Tan01}. 
The seventh set of benchmarks contains a well-studied discrete non-minimal
phase model $G_{7}$.  Non-minimal phase models cause additional
difficulties for the design of stable
controllers~\cite{Doyle:1991:FCT:574259}.
%
%
%
The eighth set of benchmarks describes the discrete model $G_{8}$ for the
\textit{Helicopter Longitudinal Motion}, which provides the longitudinal
motion dynamics of a helicopter~\cite{Franklin15}.
%
%
The ninth set of benchmarks contains the discrete model $G_{9}$ for the
known \textit{Inverted Pendulum}, which describes a pendulum dynamics with
its center of mass above its pivot point~\cite{Franklin15}.
%
%
The tenth set of benchmarks contains the \textit{Magnetic Suspension}
discrete model $G_{10}$, which describes the dynamics of a mass that
levitates with support only of a magnetic field~\cite{Franklin15}.
%
%
%
The eleventh set of benchmarks contains the \textit{Computer Tape Driver}
discrete model $G_{11}$, which describes a system to read and write data
on a storage device~\cite{Franklin15}.
%
%
The last set of benchmarks considers a discrete model $G_{12}$ that is
typically used for evaluating stability margins and controller
fragility~\cite{bhattacharyya97, keel_Bhattacharyya_examples}.

Additional benchmarks were created for the \textit{Cruise Control System},
\textit{Spring-mass damper}, and \textit{Satellite} considering parametric
additive in the nominal plant model  (represented by $\Delta_{p}\vec{G}$ in
Eq.~\eqref{digital_plant_tf}).  The uncertainties are deviations bounded to
a maximum magnitude of $0.5$ in each coefficient.  These uncertain models
are respectively represented by $G_{1b}$, $G_{2b}$, $G_{3b}$ and
$G_{3d}$.

All experiments have been conducted on a 12-core 2.40\,GHz Intel Xeon
E5-2440 with 96\,GB of RAM and Linux OS.  All times given are wall clock
times in seconds, as measured by the UNIX date command.  For the two-stage
verification engine in Figure~\ref{fig:CEGIS} we have applied a timeout of
$8$ hours per benchmark, whereas $24$ hours have been set for the approach
using a one-stage engine.

\subsection{Objectives}
\label{experimental-objectives}

Using the closed-loop control system benchmarks given in
Section~\ref{experimental-setup}, our experimental evaluation aims to answer
two research questions:
\begin{enumerate}

\item[RQ1] \textbf{(performance)} does the CEGIS approach generate a 
FWL digital controller in a reasonable amount of time?

\item[RQ2] \textbf{(sanity check)} are the synthesized controllers sound
and can their stability be confirmed outside of our model?

\end{enumerate}

\subsection{Results}
\label{experimental-results}

We give the run-times required to synthesize a stable controller for each
benchmark in Table~\ref{tab:results}.  Here, \textit{Plant} is the discrete
or continuous plant model, \textit{Benchmark} is the name of the employed
benchmark, \textit{I} and \textit{F} represent the number of integer and
fractional bits of the stable controller, respectively, 
while the two right columns display the total time (in seconds) required to synthesize a stable controller
for the given plant. 

For the majority of the benchmarks, the conjecture explained in
Section~\ref{synthesis-elements} holds and the two-stage verification engine
is able to find a stable solution in less than one minute for half of the
benchmarks.  This is possible if the inductive solutions need to be refined
with few counterexamples and increments of the fixed-point precision. 
However, the benchmark \emph{SatelliteB2} with uncertainty ($G_{3b}$) has
required too many counterexamples to refine its solution.  For this
particular case, the one-stage engine is able to complement the two-stage
approach and synthesizes a solution.  It is important to reiterate that the
one-stage verification engine does not take advantage of the inductive
conjecture inherent to CEGIS, but instead fully explores the counterexample
space in a single SAT instance.  As expected, this approach is significantly
slower on average and is only useful for benchmarks where the CEGIS approach
requires too many refinement iterations such that exploring all
counterexamples in a single SAT instance performs better.  Our results
suggest an average performance difference of at least two orders of
magnitude, leading to the one-stage engine timing out on the majority of our
benchmarks.  Table~\ref{tab:results} lists the results for both engines,
where in $16$ out of $23$ benchmarks, the two-stage engine is faster.

\begin{table}
\centering
\scalebox{0.8}{
\begin{tabular}{| r | c | l | r r || r | r |}
\hline
\# & Plant  & Benchmark                  & $I$ & $F$
   & 2-stage & 1-stage
    \\\hline\hline
1  & $G_{1a}$  & CruiseControl02
            &   4 &  16 & \textbf{12\,s}   & 67\,s     \\
2  & $G_{1b}$  & CruiseControl02$^\dagger$
            &   4 &  16 & 14600\,s   & \textbf{52\,s}     \\
3  & $G_{2a}$  & SpgMsDamper
            &  15 &  16 & \textbf{52\,s}   & 318\,s     \\
4  & $G_{2b}$  & SpgMsDamper$^\dagger$
            &  15 &  16 & \xmark  & \xmark    \\
5  & $G_{3a}$  & SatelliteB2
            &   3 &   7 & \textbf{36\,s}    & \xmark     \\
6  & $G_{3b}$  & SatelliteB2$^\dagger$
            &   3 &   7 & \xmark  & \textbf{4111\,s}  \\
7  & $G_{3c}$  & SatelliteC2
            &   3 &   5 & \textbf{3\,s}    & 205\,s    \\
8  & $G_{3d}$  & SatelliteC2$^\dagger$
            &   3 &   5 & \textbf{50\,s}  & 1315\,s    \\
9  & $G_4$  & Cruise
            &   3 &   7 & 1\,s  & 1\,s    \\
10 & $G_5$  & DCMotor
            &   3 &   7 & \textbf{1\,s}  & 10\,s    \\
11 & $G_6$  & DCServomotor
            &   4 &  11 & \textbf{46\,s}  & \xmark    \\
12 & $G_7$  & Doyleetal
            &   4 &  11 & \textbf{8769\,s}  & \xmark    \\
13 & $G_8$  & Helicopter
            &   3 &   7 & \textbf{44\,s}  & \xmark    \\
14 & $G_9$  & Pendulum
            &   3 &   7 & \textbf{1\,s}  & 14826\,s    \\
15 &$G_{10}$& Suspension
            &   3 &   7 & \textbf{1\,s}  & 5\,s    \\
16 &$G_{11}$& Tapedriver
            &   3 &   7 & 1\,s  & 1\,s    \\
17 &$G_{12a}$& a\_ST1\_IMPL1
            &  16 &   4 & \textbf{11748\,s} & \xmark   \\
18 &$G_{12a}$& a\_ST1\_IMPL2
            &  16 &   8 & \textbf{351\,s}  & \xmark   \\
19 &$G_{12a}$& a\_ST1\_IMPL3
            &  16 &  12 & \textbf{8772\,s}   & \xmark   \\
20 &$G_{12b}$& a\_ST2\_IMPL1
            &  16 &   4 & \textbf{1128\,s}  & \xmark   \\
21 &$G_{12b}$& a\_ST2\_IMPL2
            &  16 &   8 & \xmark  & \xmark    \\
22 &$G_{12b}$& a\_ST2\_IMPL3
            &  16 &  12 & \textbf{15183\,s} & \xmark   \\ 
23 &$G_{12c}$& a\_ST3\_IMPL1
            &  16 &   4 & \xmark & \xmark   \\\hline
\end{tabular}}\\[0.2ex]
\caption{\tool results ({\xmark} = time-out, $\dagger$ = 
uncertainty)
\label{tab:results}}
\end{table}

The presence of uncertainty in some particular benchmarks ($2$, $4$, $6$,
and $8$) leads to harder verification conditions to be checked by the {\sc
verify} phase, which impacts the overall synthesis time.  However,
considering the faster engine for each benchmark (marked in bold in
Table~\ref{tab:results}), the median run-time is $48$\,s, implying that
\tool can synthesize half of the controllers in less than one minute. 
Overall, the average fastest synthesis time considering both engines is
approximately $42$ minutes.  We consider these times short enough to be of
practical use to control engineers, and thus affirm RQ1.  We further observe
that the two-stage verification engine is able to synthesize stable
controllers for $19$ out of the $23$ benchmarks, and can be complemented
using the one-stage engine, which is faster for two benchmarks where the
inductive conjectures fail.  Both verification engines together enable
controller synthesis for $20$ out of $23$ benchmarks.  For the remaining
benchmarks our approach failed to synthesize a stable controller within the
time limits.  This can be addressed by either increasing either the time
limit or the fixed-point word widths considered, or by using floating-point
arithmetic instead.  The synthesized controllers have been confirmed to be
stable outside of our model representation using MATLAB, positively
answering RQ2.  A~link to the full experimental environment, including
scripts to reproduce the results, all benchmarks and the \tool tool, is
provided in the
footnote.\footnote{\url{http://www.cprover.org/DSSynth/experiment.tar.gz}\\
CBMC (SHA-1 hash) version: \\ 7a6cec1dd0eb8843559591105235f1f2c4678801}

\subsection{Threats to Validity}

We have reported a favorable assessment of \tool over a diverse set of
real-world benchmarks.  Nevertheless, this set of benchmarks is limited
within the scope of this paper and \tool's performance needs to be assessed
on a larger benchmark set in future work.

Furthermore, our approach to select suitable FWL word widths to model
plant behavior employs a heuristic based on user-provided controller
word-width specifications.  Given the encouraging results of our benchmarks,
this heuristic appears to be strong enough for the current benchmark set,
but this may not generalize.  Further experiments towards determining
suitable plant FWL configurations may thus be necessary in future work.

Finally, the experimental results obtained using \tool for stability
properties may not generalize to other properties.  The inductive nature of
the two-stage back-end of \tool increases performance significantly compared
to the one-stage back-end, but this performance benefit introduced by CEGIS 
inductive generalizations may not be observed for other controller 
properties.  Additional experiments are necessary to confirm that the
performance of our inductive synthesis approach can be leveraged in those
scenarios.

\section{Related Work}\label{sec:related}

\paragraph{Robust Synthesis of Linear Systems} 

The problem of parametric control synthesis based on stability measures for
continuous Linear Time Invariant (LTI) Single Input-Single Output (SISO)
systems has been researched for several decades.  On a theoretical level it
is a solved problem~\cite{wonham1967pole}, for which researchers
continuously seek better results for a number of aspects in addition to
stability.  A vast range of pole placement techniques such as Moore's
algorithm for eigenstructure assignment~\cite{klein1977eigenvalue} or the
more recent Linear Quadratic Regulator (LQR)~\cite{bemporad2002explicit}
have been used with increasing degrees of success.  The latter approach
highlights the importance of conserving energy during the control process,
which results in lower running costs.  Since real systems are subject to
tolerance and noise as well as the need for economy, more recent studies
focus on the problem of achieving robust stability with minimum
gain~\cite{schmid2014unified, konigorski2012pole}.  However, when applied
with the aim of synthesizing a digital controller, many of these techniques
lack the ability to produce sound or stable results because they disregard
the effects of quantization and rounding.  Recent papers on
implementations/synthesis of LTI digital controllers~\cite{das2013lqr,
ghosh2013fpga} focus on time discretization, failing to account for these
error-inducing effects and can result in digital systems that are unstable
even though they have been proven to be robustly stable in a continuous
space.

\paragraph{Formal Verification of Linear Digital Controllers} 

Various effects of discretizing dynamics, including delayed
response~\cite{Duggirala2015} and Finite Word Length (FWL)
semantics~\cite{Anta:2010:AVC:1879021.1879024} have been studied, with the
goal to either verify~\cite{daes20161} or to
optimize~\cite{oudjida2014design} given implementations.

There are two different problems that arise from FWL semantics.  The first
is the error in the dynamics caused by the inability to represent the exact
state of the physical system while the second relates to rounding errors
during computation.  In~\cite{fialho1994stability}, a stability measure
based on the error of the digital dynamics ensures that the deviation
introduced by FWL does not make the digital system unstable.  A~recent
approach~\cite{DBLP:journals/automatica/WuLCC09} uses the $\mu$-calculus to
directly model the digital controller so that the selected parameters are
stable by design.  Most work in verification focuses on finding a correct
variant of a known controller, looking for optimal parameter representations
using FWL, but ignore the effects of rounding errors due to issues of
mathematical tractability.  The analyses in~\cite{DBLP:conf/hybrid/RouxJG15,
DBLP:conf/hybrid/WangGRJF16} rely on an invariant computation on the
discrete system dynamics using Semi-Definite Programming (SDP).  While the
former uses BIBO properties to determine stability, the latter uses
Lyapunov-based quadratic invariants.  In both cases, the SDP solver uses
floating-point arithmetic and soundness is checked by bounding the error. 
An alternative approach is taken by~\cite{park2016scalable}, where the
verification of existing code is performed against a known model by
extracting an LTI model of the code through symbolic execution.  In order to
account for rounding errors, an upper bound is introduced in the
verification phase.  If~the error of the implementation is lower than this
tolerance level, then the verification is successful.

\paragraph{Robust Synthesis of FWL Digital Controllers}

There is no technique in the existing literature for automatic synthesis of
fixed-point digital controllers that considers FWL effects.

Other tools such as~\cite{economakos2016automated} are aimed at
robust stability problems, but they fail to take the FWL effects into
account.  In order to provide a correct-by-design digital controller,
\cite{alur2016compositional} requires a user-defined finite-state
abstraction to synthesize a digital controller based on high-level
specifications.  While this approach overcomes the challenges presented by
the FWL problem, it still requires error-prone user intervention. 
A~different solution that uses FWL as the starting point is an approach that
synthesizes word lengths for known control problems~\cite{jha2013swati};
however, this provides neither an optimal result nor a comprehensive
solution for the problem.

\paragraph{The CEGIS Architecture}

Program synthesis is the problem of computing correct-by-design programs
from high-level specifications, and algorithms for this problem have made
substantial progress in recent years.  One such
approach~\cite{itzhaky2010simple} inductively synthesizes invariants to
generate the desired programs.

Program synthesizers are an ideal fit for synthesis of parametric
controllers since the semantics of programs capture effects such as FWL
precisely.  In~\cite{DBLP:conf/cdc/RavanbakhshS15}, the authors use CEGIS
for the synthesis of switching controllers for stabilizing continuous-time
plants with polynomial dynamics.  The work extends to its application on
affine systems, finding its major challenge in the hardness of solving
linear arithmetic with the state-of-the-art SMT solvers.  Since this
approach uses switching states instead of linear dynamics in the digital
controller, it entirely circumvents the FWL problem.  It is also not
suitable for the kind of control we seek to synthesize.  We require a
combination of a synthesis engine with a control verification tool that
addresses the challenges presented here in the form of FWL effects and
stability measures for LTI SISO controllers.  We take the former
from~\cite{DBLP:conf/lpar/DavidKL15} and the latter from~\cite{daes20161}
while enhancing the procedure by evaluating the quantization effects of the
Hardware interfaces (ADC/DAC) to obtain an accurate discrete-time FWL
representation of the continuous dynamics.
 
\section{Conclusions}\label{sec:conclusions}

We have presented a method for synthesizing stable controllers and an
implementation in a tool called \tool.  The novelty in our approach is that
it is fully automated and algorithmically and numerically sound.  In
particular, \tool marks the first use of the CEGIS that handles plants with
uncertain models and FWL effects over the digital controller.  Implementing
this architecture requires transforming the traditional CEGIS refinement
loop into a two-stage engine: here, the first stage performs fast, but
potentially unsound fixed-point operations, whereas the second stage
restores soundness by validating the operations performed by the first stage
using interval arithmetic.  Our experimental results show that \tool is able
to synthesize stable controllers for most benchmarks within a reasonable
amount of time fully automatically.  Future work will be the extension of
this CEGIS-based approach to further classes of systems, including those
with state space.  We will also consider performance requirements while
synthesizing the digital controller.


\end{document}